\documentclass[pdflatex,sn-basic,Numbered,iicol]{sn-jnl}

\usepackage{graphicx}%
\usepackage{multirow}%
\usepackage{amsmath,amssymb,amsfonts}%
\usepackage{amsthm}%
\usepackage{mathrsfs}%
\usepackage[title]{appendix}%
\usepackage{xcolor}%
\usepackage{textcomp}%
\usepackage{manyfoot}%
\usepackage{booktabs}%
\usepackage{algorithm}%
\usepackage{algorithmicx}%
\usepackage{algpseudocode}%
\usepackage{listings}%

\theoremstyle{thmstyleone}%
\theoremstyle{thmstyletwo}%

\theoremstyle{thmstylethree}%

\newcommand{\pname}[1]{`#1'}

\begin{document}

\title[Dirigo]{\textit{Dirigo}: A Method to Extract Event Logs for Object-Centric Processes}


\author*[1]{\fnm{Jia} \sur{Wei}}\email{jia.wei@hdr.qut.edu.au}

\author*[1]{\fnm{Chun} \sur{Ouyang}}\email{c.ouyang@qut.edu.au}

\author[1]{\fnm{Arthur} \sur{ter Hofstede}}\email{a.terhofstede@qut.edu.au}

\author[2]{\fnm{Ying} \sur{Wang}}\email{ywang1@bjtu.edu.cn}

\author[2]{\fnm{Lei} \sur{Huang}}\email{lhuang@bjtu.edu.cn}

\affil*[1]{
\orgname{School of Information Systems, Queensland University of Technology}, \orgaddress{
\country{Australia}}}

\affil[2]{
\orgname{School of Economics and Management, Beijing Jiaotong University}, \orgaddress{
\country{China}}}


\abstract{
Real-world processes involve multiple object types with intricate interrelationships. Traditional event logs (in XES format), which record process execution centred around the case notion, are restricted to a single-object perspective, making it difficult to capture the behaviour of multiple objects and their interactions.  
To address this limitation, object-centric event logs (OCEL) have been introduced to capture both the objects involved in a process and their interactions with events. The object-centric event data (OCED) metamodel extends the OCEL format by further capturing dynamic object attributes and object-to-object relations. Recently OCEL 2.0 has been proposed based on OCED metamodel.
Current research on generating OCEL logs requires specific input data sources, 
and resulting log data often fails to fully conform to OCEL 2.0. Moreover, the generated OCEL logs vary across different representational formats 
and their quality remains unevaluated. 
To address these challenges, a set of quality criteria for evaluating OCEL log representations is established. Guided by these criteria, \textit{Dirigo} is proposed---a method for extracting event logs that not only conforms to OCEL 2.0 but also extends it by capturing the temporal aspect of dynamic object-to-object relations. 
Object-role Modelling (ORM), a conceptual data modelling technique, is employed to describe the artifact produced at each step of \textit{Dirigo}. To validate the applicability of \textit{Dirigo}, it is applied to a real-life use case, extracting an event log via simulation. The quality of the log representation of the extracted event log is compared to those of existing OCEL logs using the established quality criteria.
}

\keywords{Object-centric event log;
Event~log extraction;
Event~log representation;
Object role Modelling}



\maketitle

\section{Introduction}
\label{sec: intro}

Process mining has gained popularity for its ability to analyse and improve business processes in an evidence-based manner~\cite{van2023object}. Its key input---the event log---captures multidimensional time sequence data characterising process execution. Traditionally, event logs are captured in XES format~\cite{xes}, represented as tables where each row is an event capturing the execution of an activity related to a single object (the case), and each column specifies an event attribute.

Real-world processes involve multiple object types, where the relations between objects and interactions between objects and process events are complex~\cite{van2023object}. The traditional event log format is restricted to a single-object perspective and thus cannot capture the complex behaviour of real-world processes~\cite{van2023object}. To address this limitation, object-centric event logs (OCEL 1.0)~\cite{GhahfarokhiPBA21} have been introduced to 
capture the objects involved in a process and their interactions with events. The object-centric event data (OCED) meta-model~\cite{van2023object} 
further extends this by introducing dynamic object attributes and relations between objects. Recently, based on OCED, which is considered ``an early version of the OCEL 2.0 metamodel''~\cite{abs-2403-01975}, OCEL 2.0 was released.

Among the many studies~\cite{abs-2311-08795,ParkAA22,LissAA23} focused on object-centric process mining, only a few have proposed methods for generating OCEL, and these often require specific input data sources like ERP~\cite{BertiPRA21,Berti2023} and blockchain systems~\cite{Moctar-MBabaASG22}. Some work proposes methods to extract OCEL from knowledge graphs~\cite{Xiong0KMGC22,KhayatbashiHJ23} or traditional XES event logs~\cite{RebmannRA22}. Moreover, these studies rarely discuss the quality of the generated logs. Some methods do not always capture all necessary object-to-object relations and for some methods, at times, the semantics of those object-to-object relations that are captured are not entirely clear.
%

In this paper, we address existing research gaps by proposing a novel method, namely \textit{Dirigo}\footnote{\textit{Dirigo} means ``I guide'' in Latin.}, to extract event logs for object-centric processes. 
Our method is guided by the widely adopted Design Science research methodology~\cite{PeffersTRC08} and adheres to the design principles of operating at a conceptual level and maintaining generality. 
We employ Object-role Modelling (ORM)~\cite{halpin2008information}, a conceptual data modelling technique, to describe the artifact produced at each step of \textit{Dirigo}. As such, the method is not limited to specific input data sources and is independent of any particular systems, tools, or domains. The resulting OCEL log schema not only conforms to OCEL 2.0~\cite{abs-2403-01975}, but also extends it by capturing the temporal aspect of dynamic object relationships and explicitly capturing resources as event attributes. 

To develop the \textit{Dirigo} method, we first establish a set of quality criteria for evaluating OCEL log representations. 
These criteria guide the design of the \textit{Dirigo} method to ensure the generation of high-quality log representations. To validate the applicability of the proposed method, we apply it to a real-life use case, and extract an event log via simulation. 
We then assess the quality of the log representation of the extracted event log in comparison to those of existing OCEL logs, using the established quality criteria. The extracted event log from the real-life use case, along with the simulation model and evaluation experiments, are made publicly available to ensure transparency and replicability.



The remainder of this paper is structured as follows. 
Sect.~\ref{sec:preliminary} introduces key concepts in object-centric processes. 
Sect.~\ref{sec:relatedwork} reviews related work. 
Sect.~\ref{sec:methodology} outlines the research methodology, design principles, and established quality criteria for evaluation. 
Sect.~\ref{sec:dirigo} details the design of the proposed method \textit{Dirigo}. 
Sect.~\ref{sec:evaluation} presents the evaluation of the generated log using the established quality criteria. 
Sect.~\ref{sec:conclusion} summarises our findings and suggests directions for future research.


\section{Preliminaries}
\label{sec:preliminary}

The popularity of object-centric processes has increased 
due to their reflection of real-world processes \citep {GalantiLNM23}. These processes involve various objects interacting with each other rather than functioning in isolation. This section introduces the key concepts related to object-centric processes that abide by the OCEL 2.0 meta-model. 

\paragraph{\it Running example} 
%
To help illustrate these concepts, we consider 
a cargo pickup process adapted from a real-life use case at a bulk port 
in China~\citep{SongCVHW22}. 
The process starts when the customer lodges a plan for arranging trucks for pickup. 
On the scheduled date, each arranged truck arrives at the port and is weighed to record its empty weight. It then proceeds to the designated silo to load the cargo. After loading, the truck is weighed again to record the loaded weight. The port then issues a weighing ticket and a tally sheet, and the truck departs.

\paragraph{\bf Events} These represent activity occurrences in process execution~\citep{abs-2403-01975}. 
Each event has an event type (\textit{a.k.a.}\ activity), an event identifier, a timestamp, and optionally a resource. In the cargo pickup process, examples of event types include \pname{lodge a pickup plan}, 
\pname{weigh the empty truck}, \pname{load cargo}, and \pname{weigh the loaded truck}. 



\paragraph{\bf Objects} These refer to entities involved in process execution~\citep{abs-2403-01975}. They may represent physical objects like \pname{trucks} in the cargo pickup process, or business or data objects like \pname{pickup plans}. 
Each object has an object type and carries object attributes. These attributes describe various characteristics of the object and can be either static or dynamic. Static attributes (e.g., \pname{truck license plate number}) remain constant. 
In contrast, dynamic attributes (e.g., \pname{truck weight}) change due to the execution of a process event (e.g., before and after event \pname{load truck}). 

\paragraph{\bf Event-to-Object (E2O) relations} 
An event can be associated with multiple objects~\citep{van2023object}. 
This lifts the restriction of the traditional event log representation that each event must refer to a single object (i.e., the case notion). 
In the cargo pickup process, event \pname{lodge pickup plan} is associated with two object types: \pname{pickup plan} (being lodged) and \pname{cargo} (to be picked up). 
Specifying the E2O relations allows us to track the change of a particular object along the execution of the process. 
It also helps identify event attributes and their corresponding objects. 
To clarify the semantics of E2O relations, it is necessary to include a qualifier label.

\paragraph{\bf Object-to-Object (O2O) relations} 
In real-world processes, objects interact with each other, and their relations can be either static or dynamic. Static relations remain constant throughout process execution. For example, the location of stored cargo remains unchanged, establishing a static relation between \pname{cargo} and \pname{silo} (as moving the cargo is beyond the scope of this process). Conversely, dynamic relations involve changes during process execution. For example, \pname{trucks} may be reassigned to different \pname{pickup plans} upon completing their current assignments. To accurately describe the semantics of O2O relations, it is essential to include a qualifier label.

\section{Related Work}
\label{sec:relatedwork}

%
Existing methods for generating OCEL logs often require specific input data sources. 
\citet{Moctar-MBabaASG22} propose a method for extracting artifact-centric event logs (ACEL) from blockchain systems, using blockchain data to identify activities and their attributes. 
\citet{BertiPRA21,Berti2023} present a method for extracting OCEL logs from SAP ERP systems. This method identifies relevant process documents to locate relevant data tables, which are used to extract objects, object types, and O2O relations, and further to derive events and E2O relations. 

The studies of the remaining section generate event logs adhering to the OCEL~1.0 format~\citep{GhahfarokhiPBA21}, which do not capture O2O relations. 
\citet{Xiong0KMGC22} employ a virtual knowledge graph to extract OCEL logs. They suggest extracting a domain ontology from a relational database to capture relevant concept-level data, which is then mapped to OCEL concepts to extract events, objects, and E2O relations. 
\citet{KhayatbashiHJ23} propose an algorithm for transforming event knowledge graphs into OCEL logs. 
\citet{RebmannRA22} utilise traditional (XES) event logs as input to extract object-centric event data. 

%
The majority of current research presents their OCEL logs in tabular format.
ACEL \citep{Moctar-MBabaASG22} focuses on an artifact-centric log representation to capture E2O and dynamic O2O relations. 
\citet{LiMCA18} introduce an extensible object-centric (XOC) log format that captures events, E2O, and O2O relations but excludes object information to reduce complexity. 
\citet{GoossensSVA22} propose a data-aware OCEL (DOCEL) that extends OCEL~1.0~\citep{GhahfarokhiPBA21} by capturing dynamic object attributes and requiring that event attributes be linked to objects involved in process execution. 
In addition, 
\citet{BertiA19} propose a graph representation for object-centric data that includes Event-to-Object (E2O) graphs, Event-to-Event (E2E) multigraphs, and Activity-to-Activity (A2A) multigraphs. These graphs capture the relationships between events and objects in a process, the directly-following relationships between events, and the directly-following relationships between activities, respectively.

Despite various log representations having been proposed for OCEL logs, few studies have assessed their quality. 
\citet{GoossensVSSV23} identify a disconnect between object-centric logging and data-aware process modelling due to misaligned concepts. To address this, the authors propose a meta-model that unifies the terminology used in both object-centric event logs and data-centric conceptual models. The authors further compare existing log formats, focusing on whether these logs contain the terms proposed in their meta-model.
\citet{KumarST23} compare two schemas for transforming OCEL logs, the STAR schema and the normalised schema, from the perspective of relational database quality (e.g., data redundancy, integrity and information loss). However, their focus is on converting object-centric logs to relational database schemas. 

As opposed to existing research, our objective is not to introduce a new schema for object-centric logs. 
Instead, we aim to propose a method that guides the extraction of OCEL logs with high-quality log representations conforming to the OCEL 2.0 meta-model. 
Furthermore, we establish quality criteria for assessing the quality of OCEL log representations. 



\section{Approach}
\label{sec:methodology}

In this work, we aim to propose a method to extract event logs for object-centric processes. The design of this method follows the Design Science methodology and information systems development method framework discussed in Sect.~\ref{sec:instantiated_method}. Sect.~\ref{sec:qc} details the established quality criteria for evaluating object-centric event log representations, which are used to guide the development of our method.

\subsection{Research Methodology} \label{sec:instantiated_method}
Our work follows the Design Science research methodology (DSRM) proposed by~\citet{PeffersTRC08} integrated with the framework of~\citet{wijers1993modelling}. 
DSRM outlines a problem-solving process consisting of five distinct stages: identifying the problem and motivation, defining the solution objectives, design and development, demonstration and evaluation, and communication. The premise of the framework of Wij\-ers is that ``information systems development methods should include a way of thinking, controlling, modelling, working and supporting.''

%
\begin{figure*}[htbp]
\vspace*{-.75\baselineskip}
\centering
\includegraphics[width=.95\textwidth]{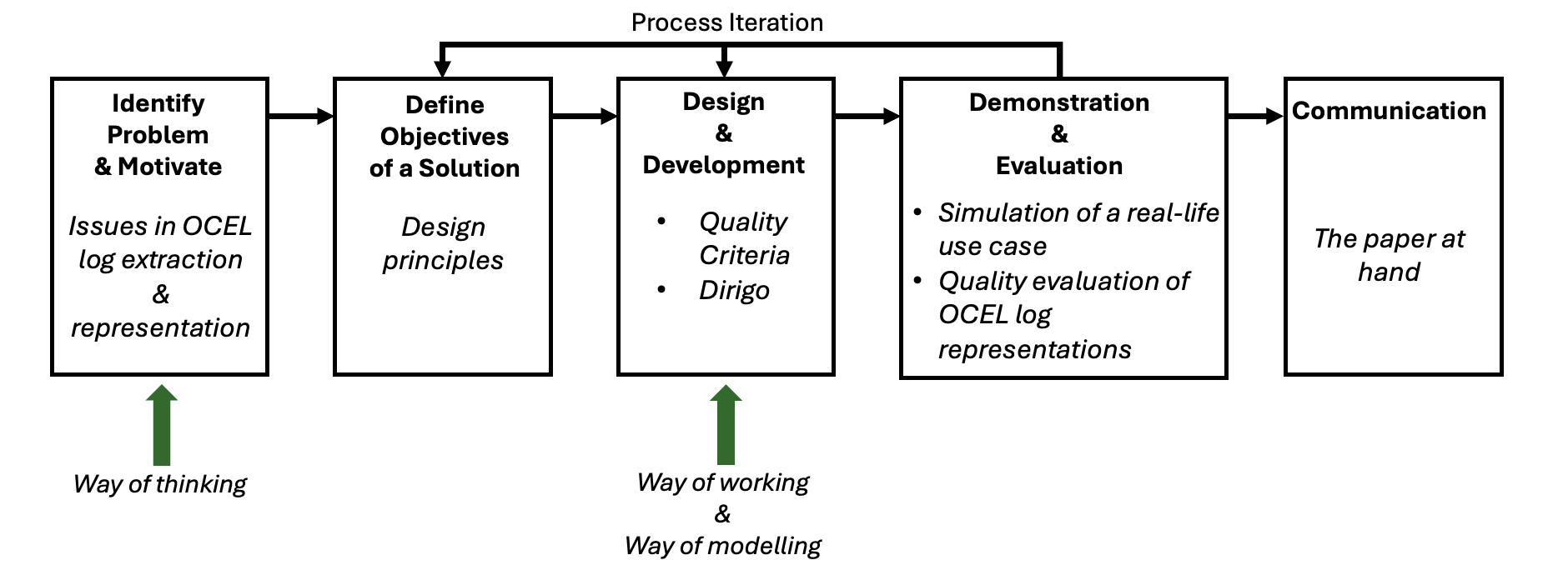}
\vspace*{-.5\baselineskip}
\caption{Overview of research methodology (instantiation of DSRM~\citep{PeffersTRC08} and integrating the framework of~\citet{wijers1993modelling})}
\label{fig:dsr}
\end{figure*}

Fig.~\ref{fig:dsr} provides an overview of our research methodology. We start 
with reviewing the literature to identify key issues in current research on OCEL log extraction and representation (discussed in  Sect.~\ref{sec:relatedwork}). This also aligns with the way of \emph{thinking} in the framework of Wijers, which addresses the `why' question, indicating the motivation and need of our study. 
Based on the identified gaps, our primary objective is to develop a method for extracting OCEL log
that adheres to the following design principles:
\begin{itemize}
    \item \textbf{Conceptualisation principle}~\citep{JARDINE19843}: The proposed method should operate at the conceptual level, the most fundamental and stable among the four information system levels (conceptual, logical, physical, and external)~\citep{halpin2008information}. This allows for independence from any specific systems and tools. 
    \item \textbf{100\% principle}~\citep{JARDINE19843}: The proposed method should ensure the capture of all information required for generating OCEL logs. 
    \item \textbf{Generic}: The proposed method should be applicable across various scenarios and independent of any specific domain.
\end{itemize}

\begin{figure*}
    \centering
    \includegraphics[width=\linewidth]{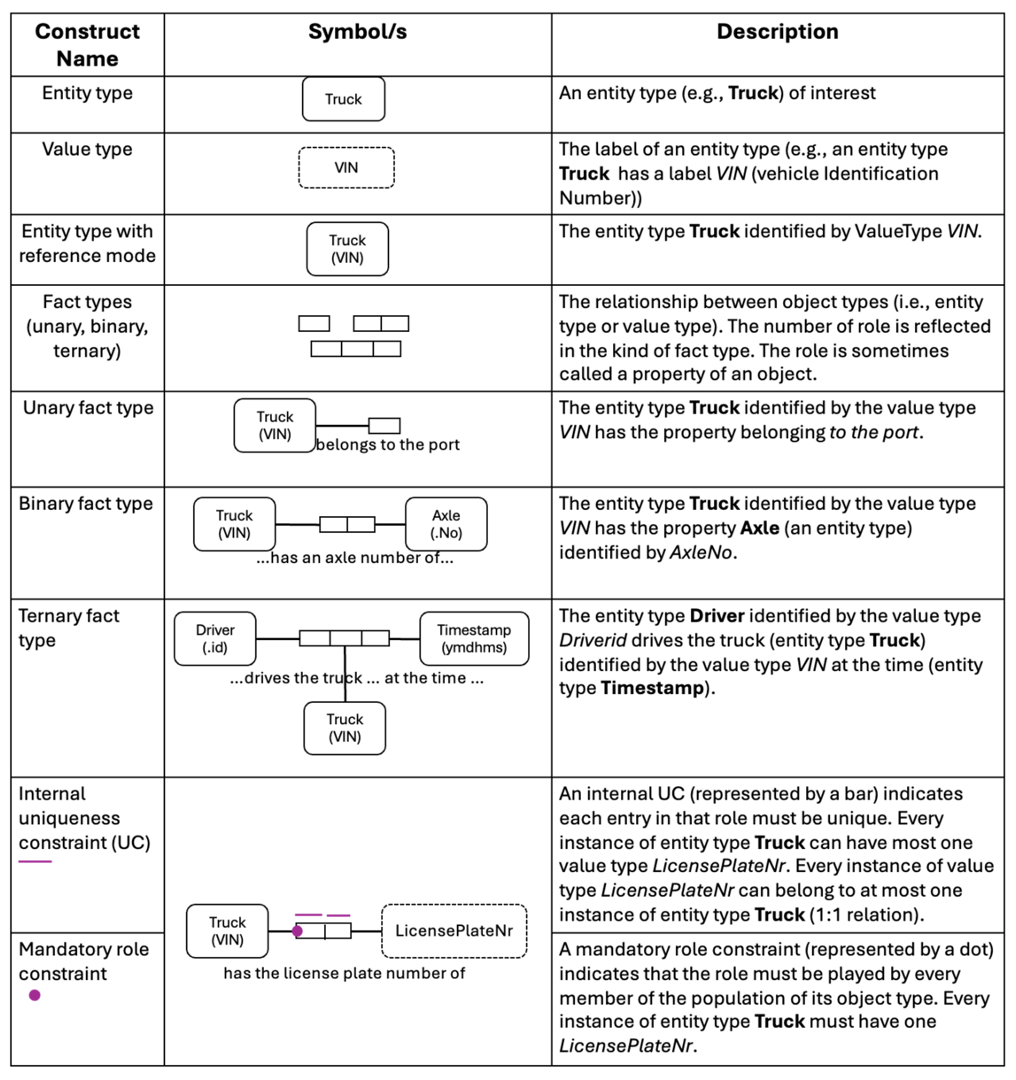}
    \caption{Subset of ORM2 Symbols (Part 1)}
    \label{fig:subsetOfORM2_1}
\end{figure*}

\begin{figure*}
    \centering
    \includegraphics[width=\linewidth]{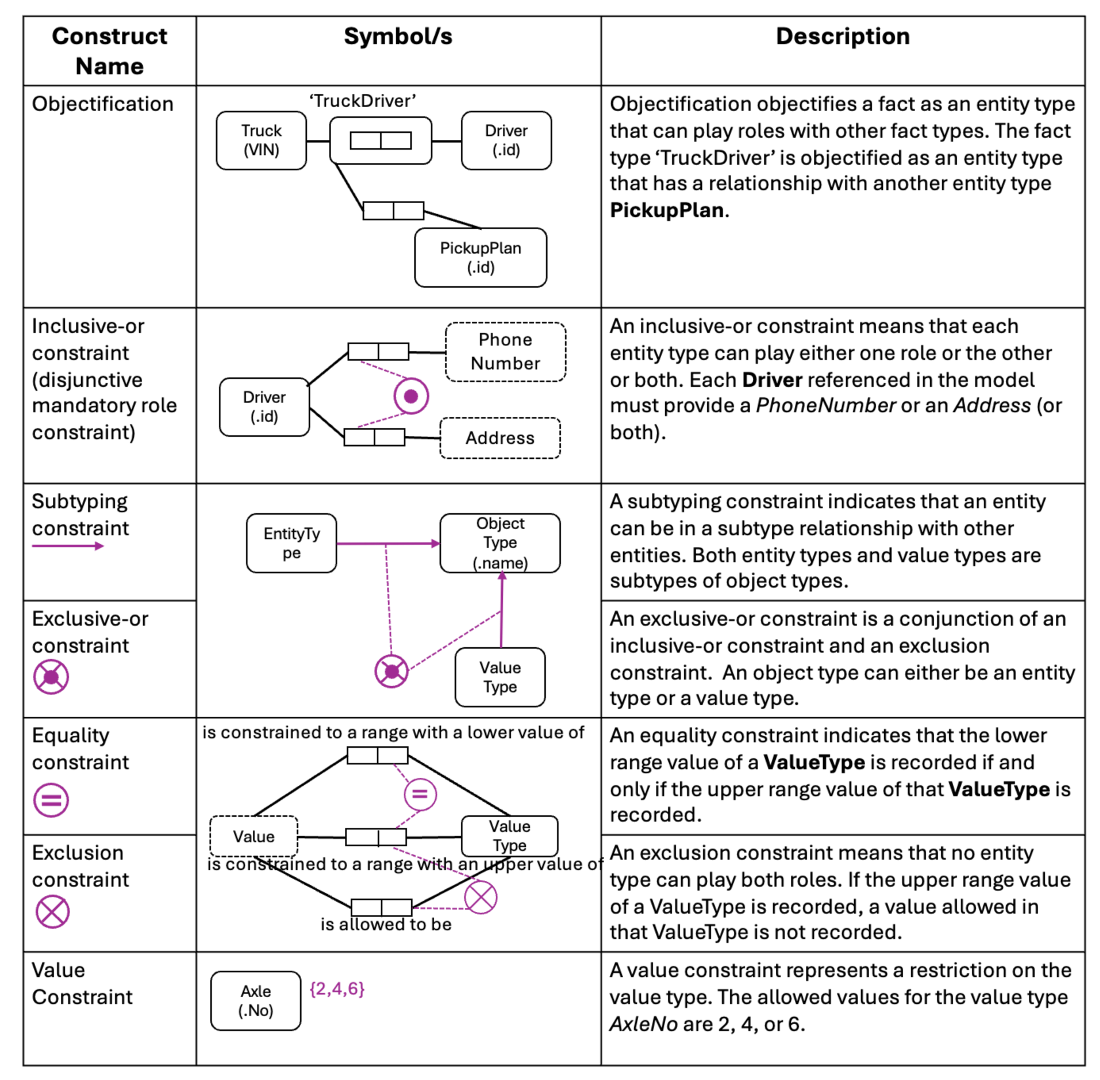}
    \caption{Subset of ORM2 Symbols (Part 2)}
    \label{fig:subsetOfORM2_2}
\end{figure*}

For design and development, we integrate the way of \emph{working} and the way of \emph{modelling} from the framework of~\citet{wijers1993modelling}.  
The way of \emph{working} is concerned with guidance for modellers towards the creation of models. Our work provides a step-by-step method for systematically obtaining information to extract OCEL logs that abide by the OCEL 2.0 meta-model~\citep{abs-2403-01975}. 
The way of \emph{modelling} is concerned with modelling concepts, their interrelationships and the rules that govern them. Fig.~\ref{fig:wayofworking} provides an overview of different levels of models involved in our method. Our work employs Object-Role Modelling (ORM)~\citep{halpin2008information}, a conceptual data modelling method, to describe the artifact produced at each step of the proposed method. 
Fig.~\ref{fig:subsetOfORM2_1} \& Fig.~\ref{fig:subsetOfORM2_2} provide the main graphical symbols in the ORM 2 notation used in this work.

We chose ORM for a number of reasons, as articulated by~\citet{halpin1999data}. Firstly, they argue that, unlike other conceptual modelling languages, ORM utilises reference schemes to identify entities, which facilitates human communication rather than relying on object identifiers. Secondly, they argue that ORM ensures greater semantic stability by modelling attributes as relationships between objects, thus avoiding changes caused by attributes in other ``data constructs''. Thirdly, Halpin and Bloesch consider ORM to ``surpass other conceptual modelling languages'' due to its use of population checking as a validation
mechanism.
%

Overall, our method involves different levels and \textit{foci}, each with their own model(s):
\begin{figure*}[t!!!]
\centering
\includegraphics[width=\textwidth]{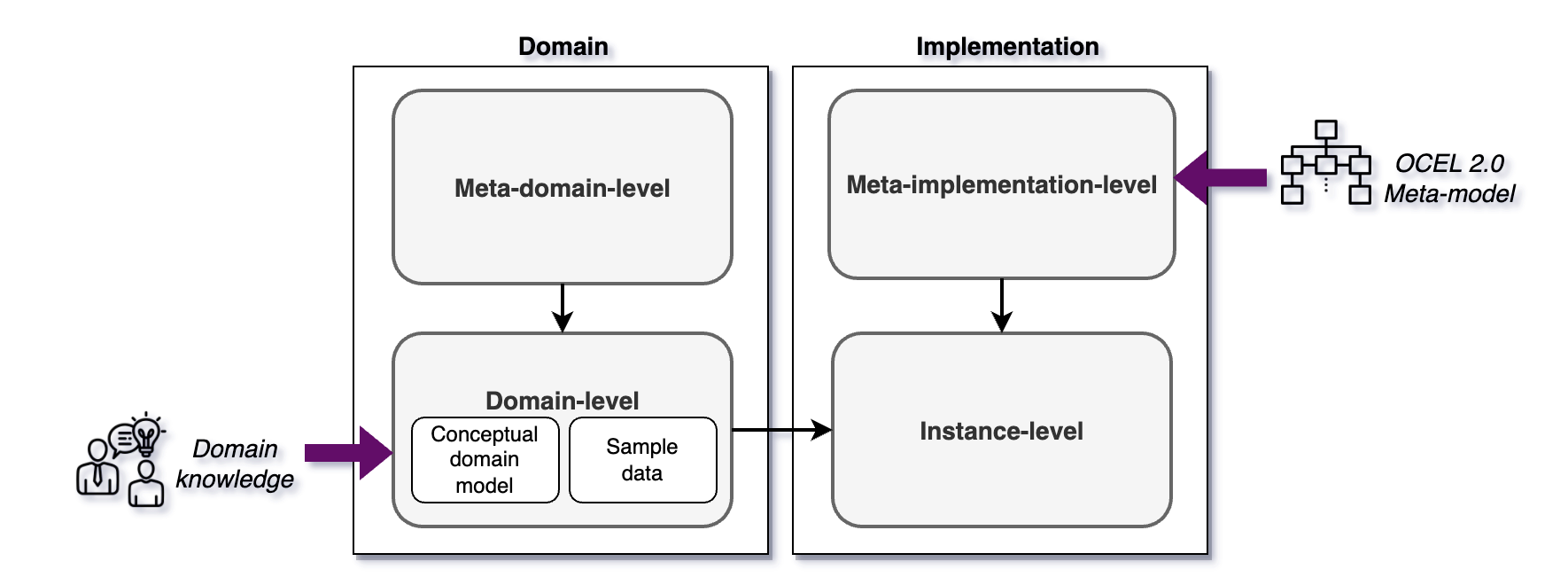}
\caption{Different levels and \textit{foci} of models in \textit{Dirigo}}
\label{fig:wayofworking}
\end{figure*}

\begin{itemize}
    \item \textbf{Meta-domain-level}: This meta level model defines the domain-focused part of our method.
           It uses a subset of ORM to capture the modelling concepts, their relationships,
           and the constraints that they need to satisfy.
    \item \textbf{Domain-level}: Domain-level models formally
           correspond to instantiations (populations) of the meta-domain-level model and, as such,
           they capture domain-specific context and information (e.g., the notion of a truck
           or a pick-up plan). We do not prescribe a specific notation for describing domain-level models, but we use, among others, ORM (with some small notational addition) and relational tables. In principle, other (data modelling) notations could also be employed (though some adaptations to the meta-domain-level model would be required). Later in this study, we will present a UML model at the domain level to illustrate this. 
           
           The domain-level models serve as a basis for the derivation
           of the implementation schema together with the meta-implementation-level model and some sample domain data. They help with the population of the meta-implementation-level model to create
           an instance-level model which prescribes the resulting object-centric schema (according to the OCEL 2.0 meta-model) to be used for the domain under consideration.
    \item \textbf{Meta-implementation-level}: This meta level model is an ORM
           representation of the OCEL 2.0 meta-model (with some small adaptation). It is used
           to ensure that the resulting OCEL log conforms to the OCEL 2.0 meta-model. 
    \item \textbf{Instance-level}: Models at this level are guided by the meta-implementation-level
           model and the domain-level model and further instantiated with sample data and,
           as such, capture actual instances of domain-specific context and information (e.g.\ not only
           names of various attributes serving as column names but also actual
           trucks or pick-up plans). The sample data can be used as a population check to determine whether the schema of the log is correct (in the way ORM uses population checking
           as a validation mechanism). Note that we do not provide an automated way of instantiating
           object-centric logs with instance-level information.
\end{itemize}

Finally, we demonstrate the proposed method using a real-life use case and evaluate the representation quality of the extracted object-centric logs using quality criteria that we establish (in the next subsection).
Note that the way of \emph{controlling}, which relates to managerial tasks, and the way of \emph{supporting}, which focuses on tool implementation for supporting information systems development in the framework of~\citet{wijers1993modelling}, are beyond the scope of our work reported in this paper. 

\subsection{Quality Criteria for Evaluation}
\label{sec:qc}

We propose quality criteria for evaluating OCEL log representation from two perspectives: 1) adherence to (some) normalisation principles of relational databases and 2) adequacy in capturing information essential for object-centric processes. 
In this subsection, we introduce and justify each criterion.

According to~\citet{KumarST23}, the normalised schema outperforms the STAR schema for transforming OCEL logs, as the latter often loses information about object relationships. Therefore, we require that all relational tables in an OCEL log should meet at least the 3rd Normal Form to ensure a high-quality representation. 
\vspace*{-.5\baselineskip}

\paragraph{\textbf{QC1 (3rd NF Compliant)}} \emph{All relational tables in an OCEL log of high-quality log representation should satisfy at least the 3rd Normal Form.}\\

According to the OCEL 2.0 meta-model, object-centric processes encompass various relationship types \citep{van2023object}. Static O2O relations remain constant throughout process execution, while dynamic O2O relations evolve with the execution of events. In addition, there are E2O relations, which link events to the objects involved. A high-quality OCEL log representation should capture all these relationships and precisely express their meaning. It should also store all attributes of objects and events related to process execution, as these attributes are essential for understanding the lifecycle of objects and the interrelationships between events. Below we detail the associated quality criteria.

\paragraph{\textbf{QC2.a (No missing static object attributes)}} \textit{A high-quality OCEL log representation should include all static attributes of objects related to process execution.}  
\vspace{-0.2cm}
\paragraph{\textbf{QC2.b (No missing dynamic object attributes)}} \emph{A high-quality OCEL log representation should include all dynamic attributes of objects related to process execution. }
\vspace{-0.2cm}
\paragraph{\textbf{QC2.c (No missing event attributes)}} \emph{A high-quality OCEL log representation should include all attributes of objects related to process execution.}
\vspace{-0.2cm}
\paragraph{\textbf{\textit{QC3.a (No missing static O2O relations)}}} \emph{A high-quality OCEL log representation should encompass all static O2O relations, i.e.\ those that remain constant during process execution.}
\vspace{-0.2cm}
\paragraph{\textbf{\textit{QC3.b (No missing dynamic O2O relations)}}} \emph{A high-quality OCEL log representation should encompass all dynamic O2O relations, i.e.\ those that change during process execution.}
\vspace{-0.2cm}
\paragraph{\textbf{\textit{QC3.c (No missing E2O relations)}}} \emph{A high-quality OCEL log representation should encompass all E2O relations.}
\vspace{-0.2cm}
\paragraph{\textbf{\textit{QC4.a (No missing labelled qualifier for O2O relations)}}} \emph{A high-quality OCEL log representation should clearly express the meaning of all O2O relations.}
\paragraph{\textbf{\textit{QC4.b (No missing labelled qualifier for E2O relations)}}} \emph{A high-quality OCEL log representation should clearly express the meaning of all E2O relations.}

\section{\textit{Dirigo}}
\label{sec:dirigo}

\begin{figure*}[t!!!]
\centering
\includegraphics[width=.85\textwidth]{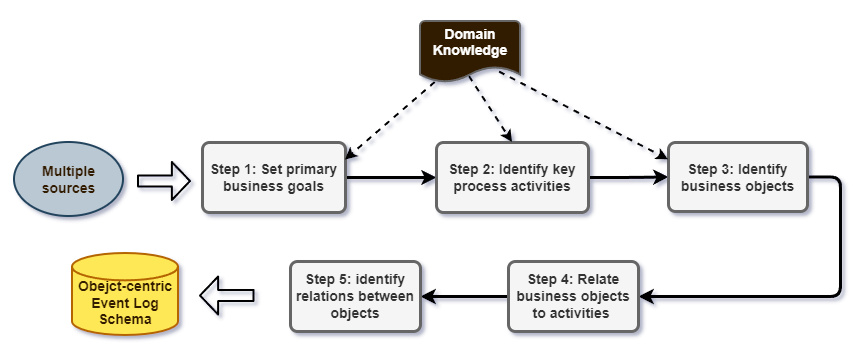}
\caption{Overview of \textit{Dirigo} and its five steps} 
\label{fig:overview}
\end{figure*}

Our method, namely \textit{Dirigo}, comprises five steps aimed at extracting event logs for object-centric processes, as illustrated in Fig.~\ref{fig:overview}. 
It begins by determining the scope of the log to be generated (Step~1), and this allows to identify key process activities (Step~2). Given the complexity of real-world processes, which often involve multiple interrelated business objects, it is crucial that the extracted event logs capture all relevant business objects (Step~3), as well as accurately reflect their relationships with events (Step~4) and interactions among objects (Step~5) during process execution. 

\subsection*{\textbf{Step 1: Set primary business goals}} 

Event logs serve as an essential data input for process mining and 
identifying the correct scope is crucial for performing analysis tasks. 
A recent survey~\citep{KampikW22} highlights that most organisations find it challenging to identify the correct scope of a process, leading to complicated process models during process discovery.  
This step aims to precisely articulate the core business objective of interest to project sponsors, specifically centred on defining the scope and purpose of event log generation. According to~\citet{JansSJ19}, business objectives typically fall into two categories: improving process efficiency or assessing regulatory compliance.
To ensure a consistent understanding of business goals among all stakeholders, it is necessary to unfold the business objectives into specific analytical questions. Examples of such questions include identifying the specific process area targeted for improvement, outlining commonly used Key Performance Indicators (KPIs), and recognising potential barriers to process execution~\citep{BenvenutiFMP22}. 

\begin{figure*}[htb]
     \centering
     \includegraphics[width=.55\textwidth]{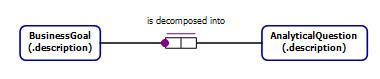}
     \caption{Meta-domain-level model of Step~1}
     \label{fig:step1}
\end{figure*}

\paragraph{\textbf{Meta-domain-level model} (Fig.~\ref{fig:step1})} The meta-domain-level model for Step~1 -- specified using ORM -- captures the relationship between business goals and associated analytical questions. It comprises two entity types: \textit{BusinessGoal} and \textit{AnalyticalQuestion}, identified by descriptions.

\subsection*{\textbf{Step 2: Identify key process activities}} 
In light of the business objectives set in Step~1, this step focuses on the identification of the key activities within the process guided by domain knowledge. Notably, candidate activities are expected to possess time-related information stored in the database~\citep{JansSJ19}. In particular, the following questions need to be addressed: 
\begin{enumerate}
    \item Where does the process begin and end?
    \item What critical activities are vital for the execution of the process?
    \item Among these critical activities, which ones should be modelled to address the analytical questions identified in Step~1?
    \item What resources (human or systems) are involved in these activities?
\end{enumerate}

\begin{figure*}[t!]
\centering
\includegraphics[width=.95\textwidth]{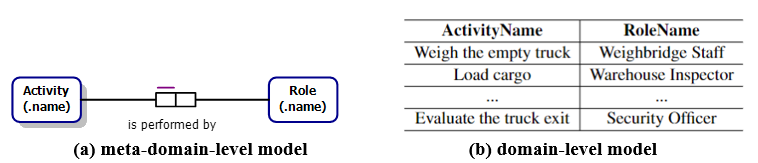}
\caption{Meta-domain-level and an example of partial domain-level models of Step~2}
\label{fig:step2}
\end{figure*}

\paragraph{\textbf{Meta-domain-level model} (Fig.~\ref{fig:step2}a)} The meta-domain-level model specified in ORM captures the relationships between process activities and their corresponding resources. It comprises two entity types: \textit{Activity} and \textit{Role}, identified by their names.

\paragraph{\textbf{Domain-level model} (Fig.~\ref{fig:step2}b)} This captures an instantiation of the above meta-domain-level model in the context of the port use case. 
The cargo pickup process consists of activities such as \pname{Lodge pickup plan} at the start and \pname{Evaluate the truck exit} at the end. 
Truck weighing activities are monitored by \pname{weighbridge staff} while \pname{warehouse inspectors} supervise cargo loading. 
When the trucks leave the port, they are assessed by a \pname{security officer}. This domain-level model is presented in a relational table format.

\begin{figure*}[h!]
\centering
\includegraphics[width=.65\textwidth]{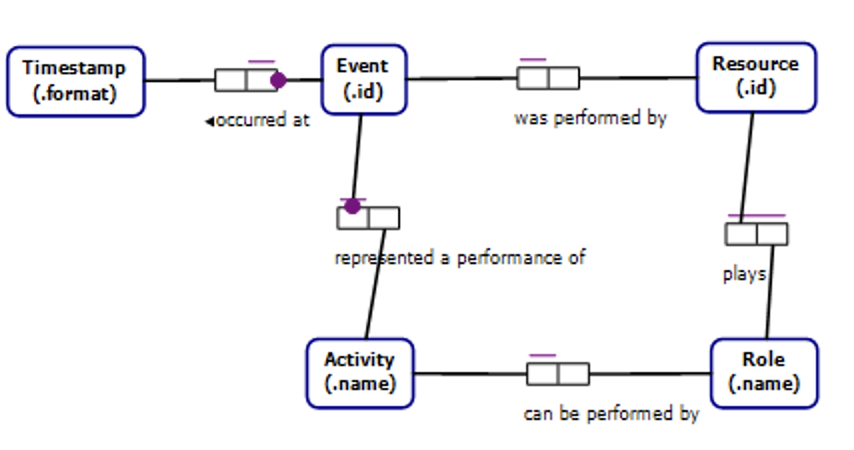}
\caption{Meta-implementation-level model of Step~2} 
\label{fig:step2_bridging}
\end{figure*}

\paragraph{\textbf{Meta-implementation-level model} (Fig.~\ref{fig:step2_bridging})} This meta-implementation-level model, specified in ORM, encapsulates the components of the OCEL 2.0 meta-model related to \textbf{Events}. Each event is uniquely identified by an event ID, includes a single activity (event type), and is associated with a timestamp. Though the resource of an event can be treated as another attribute, we make this involvement explicit in our model (the OCEL 2.0 meta-model does not explicitly mention resources). 
Note that we leave the timestamp format unspecified as it may have to be adapted on a case-by-case basis. For timestamps in the models of the subsequent steps, we apply the same approach.


\begin{figure*}[h!]
\centering
\includegraphics[width=.85\textwidth]{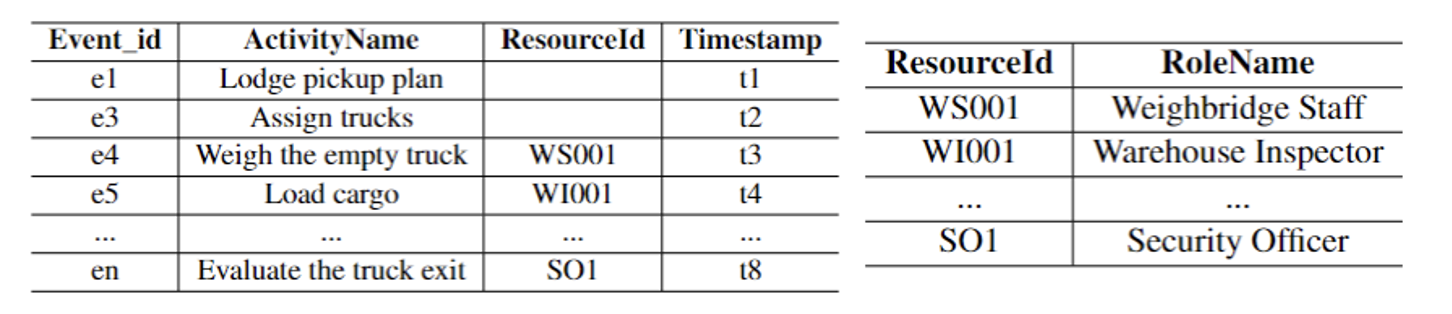}
\caption{An example of the instance-level model of Step~2} 
\label{fig:step2_instance}
\end{figure*}

\paragraph{\textbf{Instance-level model} (Fig.~\ref{fig:step2_instance})} Each event is uniquely identified and timestamped, and in this model, the \texttt{ActivityName} and \texttt{ResourceId} columns are populated from the domain-level model. For instance, an event with \texttt{Event\_id} `e4' is associated with activity \pname{Weigh the empty truck} and performed by weighbridge staff \pname{WS001} at timestamp \pname{t3}.
\pname{WS001} is an instance of \texttt{ResourceId}. These instances are part of the domain though not represented in the domain-level model where only the roles they can play are modeled. A sample of these resources needs to be added to the instance-level model serving as a population check. Those that are added need to be linked to the \texttt{Role}(s) they play in the domain. As this relationship is many-to-many, this link is captured in a separate table to avoid redundancy.

\subsection*{\textbf{Step 3: Identify business objects}}

Real-life processes entail intricate scenarios where events are linked to multiple object types, leading to complex interrelations among them. 
This step is dedicated to identifying all relevant objects within the process guided by domain knowledge. For example, the cargo pickup process involves various object types, such as trucks, cargoes, and pickup plans, and a single pickup plan may be associated with different cargoes. 

\begin{figure*}[h!]
\centering
\includegraphics[width=\textwidth]{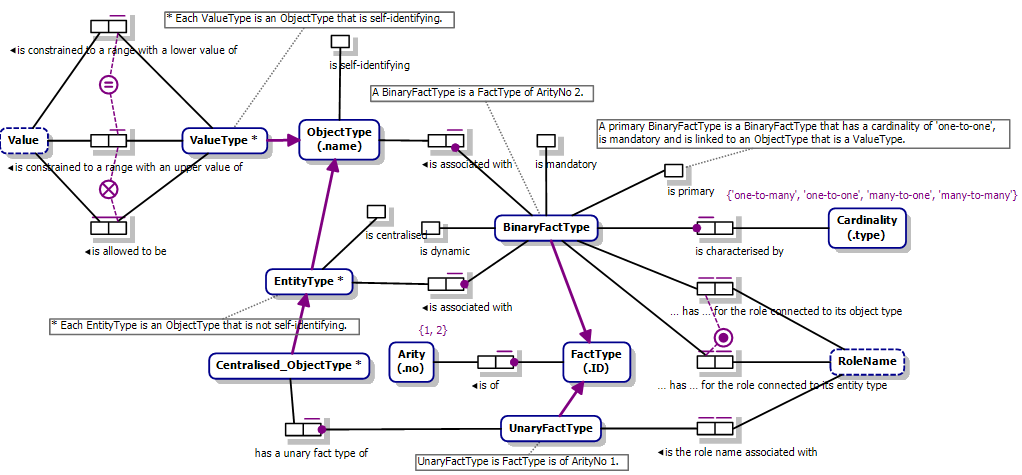}
\caption{Meta-domain-level model of Step~3} 
\label{fig:step3}
\end{figure*}

\paragraph{\textbf{Meta-domain-level model} (Fig.~\ref{fig:step3})} The meta-domain-model specified using ORM at Step~3 prescribes the concepts necessary to capture objects and corresponding attributes\footnote{Our meta-domain-model does not directly have attributes. As in ORM, these are captured through unary or binary fact types.}, which further provides guidance for developing domain-level models. The central element in the meta-domain-level model is \textit{ObjectType} which encapsulates things of interest. 

\begin{figure*}[h!]
\centering
\includegraphics[width=\textwidth]{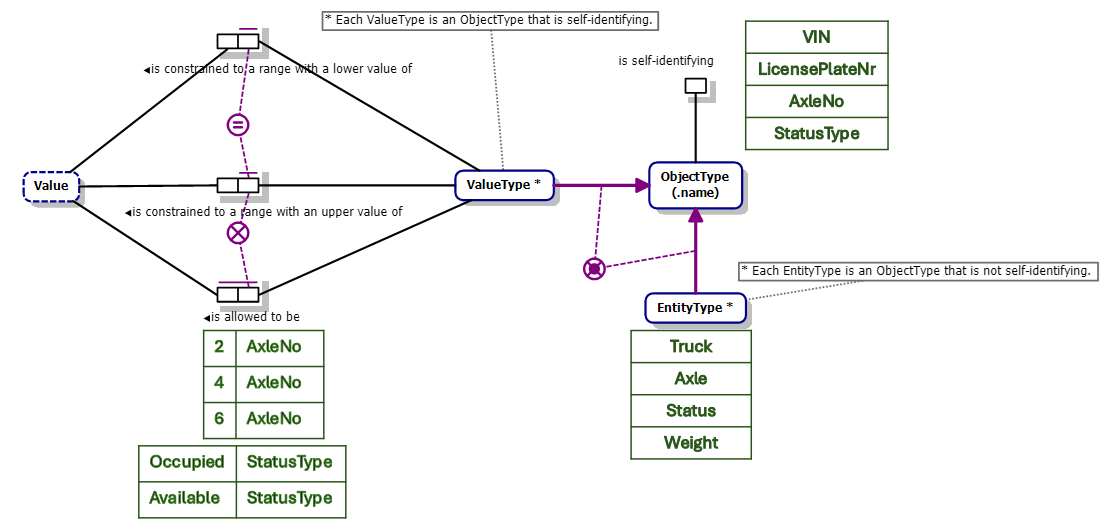}
\caption{Population of part of the meta-domain-level models for Step~3 (Fig.~\ref{fig:step3})} 
\label{fig:step3a_population}
\end{figure*}

\begin{figure*}[h!]
\centering
\includegraphics[width=\textwidth]{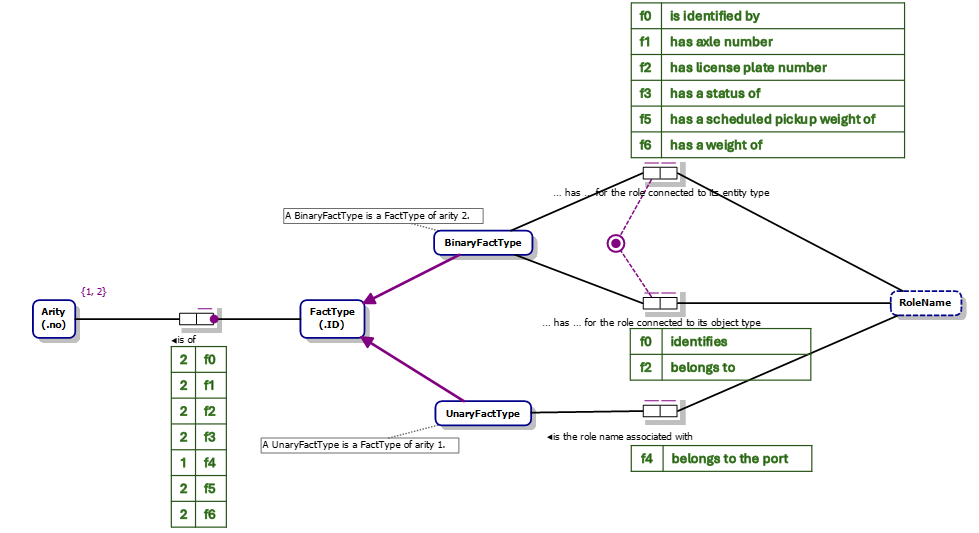}
\caption{Population of part of the meta-domain-level models for Step~3 (Fig.~\ref{fig:step3})} 
\label{fig:step3b_population}
\end{figure*}

\begin{figure*}[h!]
\centering
\includegraphics[width=.85\textwidth]{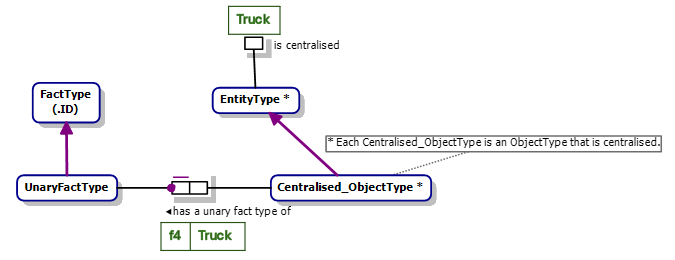}
\caption{Population of part of the meta-domain-level models for Step~3 (Fig.~\ref{fig:step3})} 
\label{fig:step3c_population}
\end{figure*}

\begin{figure*}[h!]
\centering
\includegraphics[width=\textwidth]{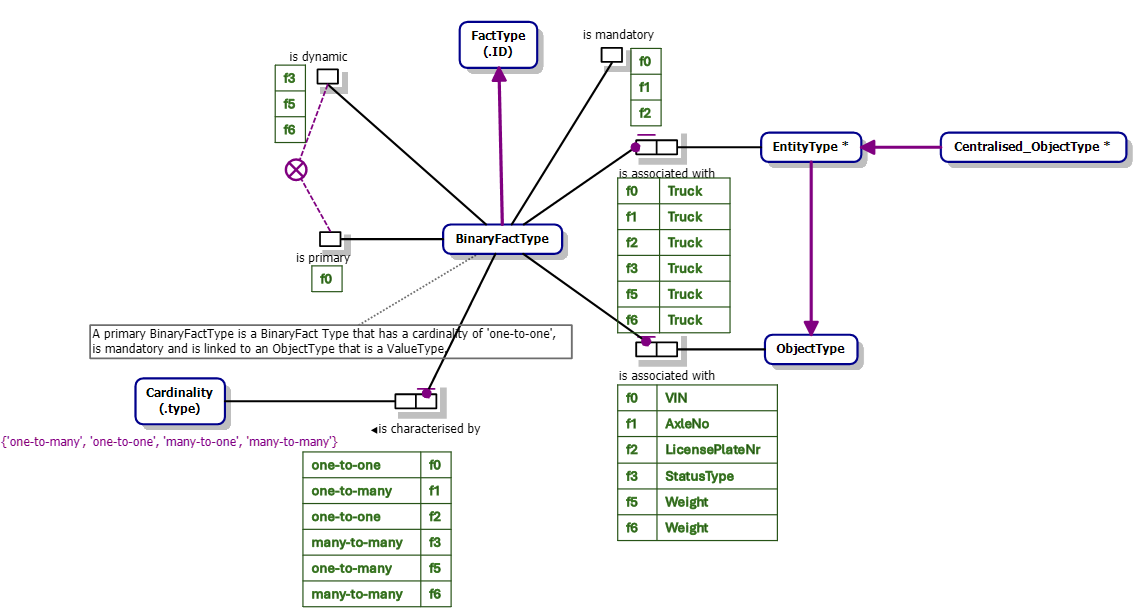}
\caption{Population of part of the meta-domain-level models for Step~3 (Fig.~\ref{fig:step3})} 
\label{fig:step3d_population}
\end{figure*}

\begin{figure*}[h!]
\centering
\includegraphics[width=\textwidth]{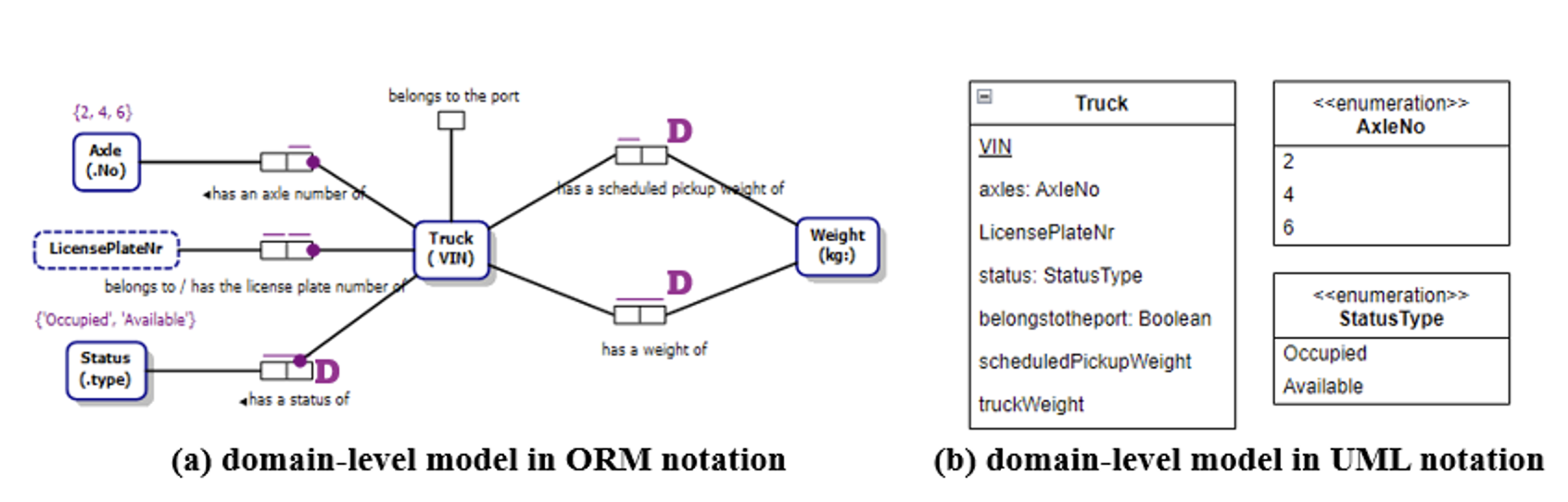}
\caption{Partial domain-level models of Step~3 in the cargo pickup process} 
\label{fig:step3_domain}
\end{figure*}

{\paragraph{\textbf{Domain-level model} (Fig.~\ref{fig:step3_domain})}

In the cargo pickup process, examples of \textit{ObjectType}s include~\pname{AxlesNo}, \pname{LicensePlateNr},~\pname{Truck},~\pname{Weight}. The \textit{ObjectType} can take its values from a given set or from a range of values. For instance, the \pname{AxlesNo} can carry values of 2, 4 or 6. 
These objects are modelled either as \textit{EntityType} or \textit{ValueType}. 
Each \textit{ValueType} is an object that is self-identifying as its constant value is understood within a given context, such as \pname{LicensePlateNr}. Conversely, \textit{EntityType} is not self-identifying and relies on a reference mode for identification. E.g., each \pname{Truck} can be identified by a unique vehicle identification number \pname{VIN}. This reflects a segment of the meta-domain-level model with its population detailed in Fig.~\ref{fig:step3a_population}.

In our meta-domain-level model, we only include unary and binary fact types, and each fact type is identified by its id. In our model, \textit{BinaryFactType}\footnote{Rather than write \emph{BinaryFactTypeName}, we simply write \emph{BinaryFactType} and thus drop the underlying value type as this is more convenient. We will also do this for other entity types.} refers to relations between an \textit{EntityType} and an \textit{ObjectType} (ORM allows for a more general notion of binary fact types). Some binary fact types are not explicitly shown, namely, those that serve as the identifying relationship for some object type. In this example, binary fact type `f0' is associated with \textit{EntityType} \pname{Truck} and \textit{ObjectType} \pname{VIN}, and the relationship description is \pname{is identified by}, attached to the role played by \pname{Truck} in this fact type. Thus, `f0' is primary and mandatory as a \pname{VIN} is the reference mode for identifying instances of object type \pname{Truck}. Moreover, \pname{VIN} is unique to each truck; thus, the cardinality between these two object types is a one-to-one relationship. Each \textit{EntityType} is associated with attributes distinct from the object's reference mode. A \pname{Truck} has attributes such as different types of weight. For example, each \pname{Truck} possesses a scheduled pickup weight, an empty weight before loading the cargo, and a loaded weight. Consequently, it becomes imperative to employ \textit{RoleName} for distinguishing attributes pertaining to truck weight. Fig.~\ref{fig:step3b_population} and Fig.~\ref{fig:step3d_population} provide a sample population of part of the model concerning fact types.

Moreover, these attributes exhibit static or dynamic characteristics, and some may be mandatory while others remain optional. For instance, `f1' is a mandatory binary fact type indicating that each truck has a mandatory static attribute, \pname{AxleNo}, containing a value that remains constant for each individual truck.  Conversely, \pname{Truck weight} is considered an optional and dynamic attribute, contingent upon the unfolding of specific events. In this particular scenario, the execution of the event \pname{Weigh the empty truck} leads to a value for the attribute that captures the empty weight of the truck. Subsequently, upon executing the event \pname{Load the cargo}, the truck acquires a value for the attribute that captures its loaded truck weight. As a truck may have multiple weights, `f6' is a many-to-many relationship. This reflects the meta-domain-level model in Fig.~\ref{fig:step3}, and Fig.~\ref{fig:step3d_population} provides a sample population of this model, which constitutes part of a domain-level model. 

We further propose a sub-type of \textit{EntityType}, namely \textit{Centralised\_ObjectType}, which captures all relevant business objects within the process (Fig.~\ref{fig:step3}). In the cargo pickup process, \pname{Truck}, \pname{Cargo} and \pname{Pickup plan} are classified as \textit{Centralised\_ObjectType}. A \textit{Centralised\_ObjectType} may play a role in a \textit{UnaryFactType}. Unary fact types capture boolean attributes of an object, such as the \pname{belongs to the port} attribute of a truck. The population of this segment of the meta-domain-level model is shown in Fig.~\ref{fig:step3c_population}. The aggregation of all previously described populations produces a domain-level model which can be depicted in ORM notation, as shown in Fig.~\ref{fig:step3_domain}a. The capital \textbf{D} used in the ORM model is our ad hoc notation to indicate dynamic attributes of an object type (this is not part of the ORM notation). In principle, another concrete representation can be chosen for this domain-level model. To illustrate this, we show this domain-level model in the UML notation (Fig.~\ref{fig:step3_domain}b), which was derived through a mapping from the ORM model~\citep{halpin2008information}. Note though that this UML model is no longer an exact instance of our meta-model, so some adaptation is needed to fully support the use of UML as a concrete representation for domain-models produced by this step. 


\paragraph{\textbf{Meta-implementation-level model} (Fig.~\ref{fig:step3_bridging})} This meta-implementation-level model specified in ORM encapsulates the components of the OCEL 2.0 meta-model related to \textbf{Objects}. Each object is uniquely identified by an object id. The model captures value changes of attributes of objects at particular moments of time (as recorded in the OCEL log). 

\begin{figure*}[h!]
\centering
\includegraphics[width=.55\textwidth]{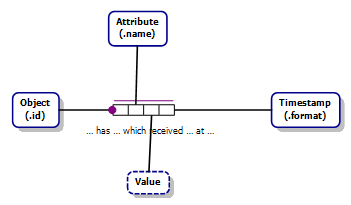}
\caption{Meta-implementation-level model of Step 3} 
\label{fig:step3_bridging}
\end{figure*}

\paragraph{\textbf{Instance-level model} (Table~\ref{tab:instance-level-step3})} In this model, the \texttt{Object\_Id} column is populated by instances of the value type linked to the centralised object type through a primary binary fact type. These values thus identify the objects. The \texttt{AttributeName} column is populated by other binary or unary fact types which describe attributes of centralised object types, derived from the domain-level model. For instance, a specific instance of \textit{Centralised\_ObjectType} \pname{Truck}, identified by the object\_id \pname{Tr1}, is characterised by having `4' axles, license plate number \pname{LPT1}, and status \pname{Available} at timestamp \pname{t0}. In addition, it possesses a scheduled pickup weight denoted as \pname{schw1} recorded at timestamp \pname{t2}. This particular truck instance \pname{Tr1} has an empty truck weight of \pname{emptw} recorded at timestamp \pname{t3} and its loaded truck weight of \pname{loadedw} recorded at timestamp \pname{t5}.

\begin{table}[]
\centering
\resizebox{.5\textwidth}{!}{%
\begin{tabular}{c|c|c|c}
\hline
\textbf{Object\_id} & \textbf{Timestamp} & \textbf{Attribute}             & \textbf{Value}                 \\ \hline
Tr1                 & t0                 & AxleNo                & 4                              \\ \hline
Tr1                 & t0                 & LicensePlateNr        & LPT1                           \\ \hline
Tr1                 & t0                 & TruckStatus           & \multicolumn{1}{l}{Available} \\ \hline
Tr1                 & t2                 & ScheduledPickupWeight & schw1                          \\ \hline
Tr1                 & t2                 & TruckStatus           & Occupied                       \\ \hline
Tr1                 & t3                 & TruckWeight           & emptw                          \\ \hline
Tr1                 & t5                 & TruckWeight          & loadedw                        \\ \hline
\end{tabular}%
}
\caption{Partial instance-level model of object type \pname{Truck}}
\label{tab:instance-level-step3}
\end{table}


\subsection*{\textbf{Step 4: Relate business objects to activities}}
Real-world processes often involve multiple objects associated with a single event. For instance, in the cargo pickup process, the event \pname{Assign trucks} entails two objects: \pname{Pickup Plan} and \pname{Truck}. This arises because, after a company lodges a pickup plan, they need to assign trucks to pick up the cargo based on the scheduled total pickup weight.

\paragraph{\textbf{Meta-domain-level model} (Fig.~\ref{fig:step4}a)} This meta-domain-level model, specified in ORM, captures the relationship between activities and centralised object types, as well as the meaning of their relationships.

\paragraph{\textbf{Domain-level model} (Fig.~\ref{fig:step4}b)} This domain-level model, presented in a tabular format, captures the relationship between activities and associated centralised object types involved in the cargo pickup process. For example, activity \pname{Lodge pickup plan} is associated with centralised object types \pname{Pickup plan} and \pname{Cargo}.

\begin{figure*}[h]
\centering
\includegraphics[width=\textwidth]{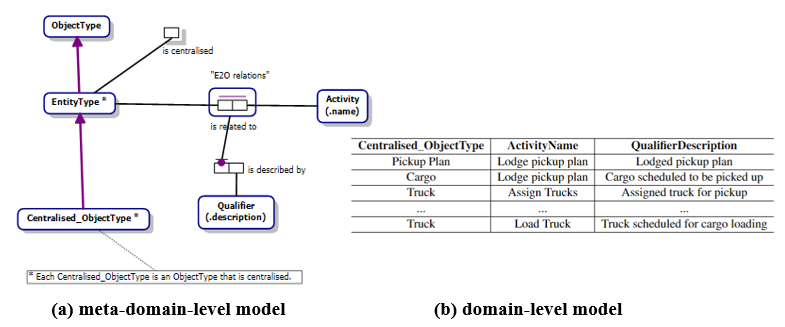}
\caption{Meta-domain-level and an example of partial domain-level model for Step 4} \label{fig:step4}
\end{figure*}

\paragraph{\textbf{Meta-implementation-level model} (Fig.~\ref{fig:step4_bridging})}
The meta-implementation-level model for Step~4, specified in ORM, captures the Event-to-Object Relations introduced in the OCEL 2.0 meta-model. It includes concepts such as events, associated objects, and qualifiers for capturing the meaning of their relationships.

\begin{figure*}[h]
\centering
\includegraphics[width=.45\textwidth]{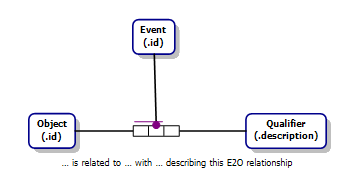}
\caption{Meta-implementation-level model of Step 4} \label{fig:step4_bridging}
\end{figure*}

\paragraph{\textbf{Instance-level model} (Table~\ref{tab:instance-level-step4})}  This model glues together objects and events (instances of \texttt{Event\_id} as specified in the meta-implementation-level model) by qualifying the way the former participate in the latter. For instance, event `e1' (\pname{Lodge pickup plan}) is associated with two object instances: pickup plan \pname{Pcp1} and cargo \pname{Cid}. According to the \texttt{E20\_Qualifier}, we can derive that the pickup plan was lodged, and the cargo was scheduled to be picked up when `e1' was executed. Objects are instances of the type Object in the meta-implementation-level model and correspond to instances of the centralised object type in the domain-level model.


\begin{table}[]
\centering
\resizebox{.5\textwidth}{!}{%
\begin{tabular}{c|c|c}
\hline
\textbf{Event\_id} & \textbf{Object\_id}                         & \textbf{E2O\_Qualifier} \\ \hline
e1                 & Pcp1                 &  Lodged pickup plan \\ \hline
e1                 & Cid1                  &  Cargo scheduled to be picked up                        \\ \hline
e2                 & Tr1                  & Assigned truck for pickup                      \\ \hline
 ... & ... & ... \\ \hline
e5                 & Tr1                    & Truck scheduled for cargo loading                         \\ \hline
\end{tabular}%
}
\caption{Partial instance-level model of Step 4 in the scenario of the cargo pickup process}
\label{tab:instance-level-step4}
\end{table}


\subsection*{\textbf{Step 5: Identify relationships among objects}}
This step emphasises capturing the intricate interrelations among the business objects identified in Step~3 effectively. 

\paragraph{\textbf{Meta-domain-level model} (Fig.~\ref{fig:step5}a)}  The meta-domain-level model specified in ORM captures the relationships between centralised object types and their meaning.


\paragraph{\textbf{Domain-level model} (Fig.~\ref{fig:step5}b)} A pickup plan schedules multiple trucks, each assigned to collect a portion of specific bulk cargo at designated times. Consequently, a centralised object type, \pname{Pickup Plan}, is associated with another centralised object type, \pname{Truck}, and this relationship may change during process executions. For example, a truck is assigned to a pickup plan and, upon finishing the pickup, can be dropped from this pickup plan. To differentiate this dynamic association between trucks and pickup plans, it is critical to specify the qualifier description. A sample population of the meta-domain-level model for this step is depicted in Fig.~\ref{fig:step5}b. 

\begin{figure*}[t!]
\centering
\includegraphics[width=\textwidth]{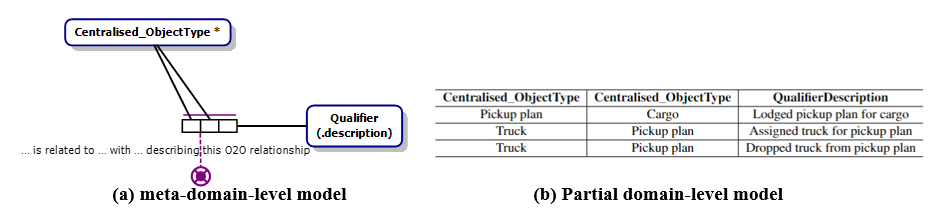}
\caption{Meta-domain-level and an example of partial domain-level models for Step 5} 
\label{fig:step5}
\end{figure*}

\paragraph{\textbf{Meta-implementation-level model} (Fig.~\ref{fig:step5_bridging})} This meta-implementation-level model, specified in ORM, captures the Object-to-Object Relations introduced in the OCEl 2.0 meta-model. It includes concepts such as (source) objects, (target) objects, and qualifiers to describe the meanings of their relationships. Further, we include the timestamp of the O2O relations; by mapping the timestamp of the corresponding event, we can derive when these dynamic O2O relationships occur.
\begin{figure*}[h]
\centering
\includegraphics[width=.45\textwidth]{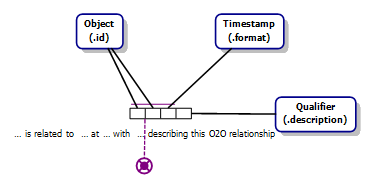}
\caption{Meta-implementation-level model of Step 5} \label{fig:step5_bridging}
\end{figure*}

\paragraph{\textbf{Instance-level model} (Table~\ref{tab:instance-level-step5})} In this model, the instances in column \texttt{Object\_id} are instances of type Object in the meta-implementation-level model and these instances all are also instances of the centralised object types identified in the domain-level model. Based on the E2O relationships identified in Step~4, we can understand the relationship between events and objects. When multiple objects are related to the same event this needs to be reflected in the O2O relationship table and a corresponding timestamp needs to be recorded.
This we can achieve by using the E2O relationships identified in Step~4 to determine in which events these O2O relationships occur. We populate the \texttt{Timestamp} column with data from the instance-level model in Step~4. For example, the relationship between objects \pname{Tr1} and \pname{Pcp1} changes from timestamp \pname{t2} to \pname{t8}. From the recorded time, we can determine that the truck with object\_id \pname{Tr1} was assigned to pickup plan \pname{Pcp1} when the event \pname{Assign trucks} occurred and dropped from pickup plan \pname{Pcp1} when the event \pname{Evaluate the truck exit} occurred.

\begin{table*}[]
\centering
\resizebox{.75\textwidth}{!}{%
\begin{tabular}{c|c|c|c}
\hline
\textbf{(source) Object\_id} & \textbf{(target) Object\_id} & \textbf{Timestamp} & \textbf{o2o\_qualifier}        \\ \hline
Pcp1                        & Cid1                        & t1                 & Lodged Pickup Plan for Cargo   \\ \hline
Pcp3                        & Cid4                        & t3                 & Lodged Pickup Plan for Cargo   \\ \hline
Tr1                         & Pcp1                        & t2                 & Assigned Truck for Pickup Plan \\ \hline
Tr1                         & Pcp1                        & t8                 & Dropped Truck from Pickup Plan \\ \hline
\end{tabular}%
}
\caption{Partial Instance-level model of Step 5}
\label{tab:instance-level-step5}
\end{table*}


At this stage, with all the instance-level models available, the focus shifts to integrating this information into the desired event log schema. For instance, the resulting object-centric log schema (in Fig.~\ref{fig:dirigo_schema}) includes four types of relational tables that abide by the OCEL 2.0 meta-model:
\begin{itemize}
    \item \textbf{\textit{Events} table} (Fig.~\ref{fig:dirigo_schema}b): Contains all event information, including event\_id, activity\_name, timestamp and resources (if applicable). This table constitutes the instance-level model for Step~2.
    \item \textbf{\textit{Object} tables} (Fig.~\ref{fig:dirigo_schema}d \& Fig.~\ref{fig:dirigo_schema}e): A group of tables, each capturing information related to a specific type of object, including object\_id, timestamp, static attributes, and dynamic attributes. For each object type table, we transform the instance-level model from Step~3. Specifically, we relocate attribute names to the \texttt{ocel\_changed\_field} column (this column name is also used in some sample logs on the OCEL-standard web site\footnote{\url{https://ocel-standard.org/}}). When an entry records values for a collection of static attributes, the corresponding \texttt{ocel\_changed\_field} has no value, and the timestamp for this entry indicates when these values were recorded. When the name of a dynamic attribute appears as an entry in the \texttt{ocel\_changed\_field} column, this means that its value changed at the timestamp recorded and to the value for that attribute recorded in that row (which can be found in the column with the name of the dynamic attribute).
    \item \textbf{\textit{E2O relation} table} (Fig.~\ref{fig:dirigo_schema}a): Captures relationships between events and objects, including the event\_id, object\_id and their relationship qualifiers. This table constitutes the instance-level model for Step~4.
    \item \textbf{\textit{O2O relation} table} (Fig.~\ref{fig:dirigo_schema}c): Captures dynamic relationships between object\_ids, the time when this dynamic O2O relationship occurred, and the corresponding relationship qualifier. This table constitutes the instance-level model for Step~5.
\end{itemize}

\begin{figure*}[h]
\centering
\includegraphics[width=\textwidth]{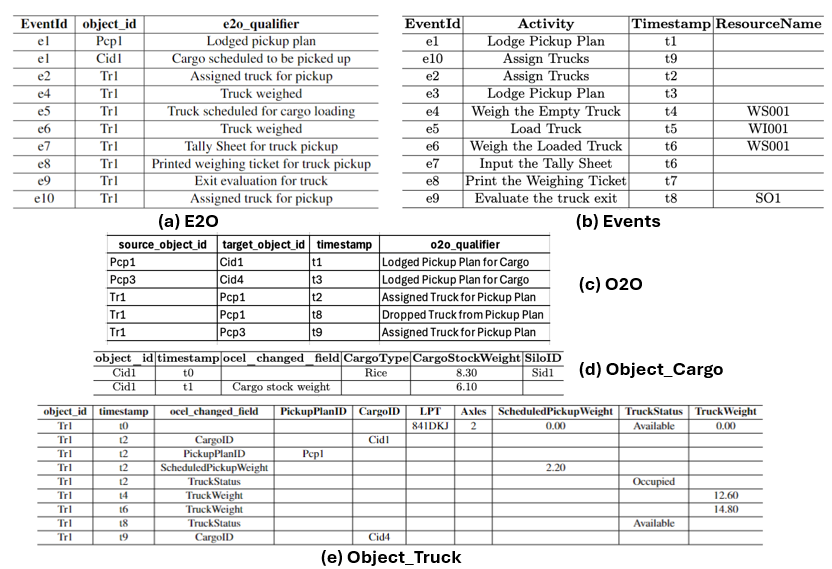}
\caption{The OCEL log schema extracted following \textit{Dirigo}} \label{fig:dirigo_schema}
\end{figure*}

To illustrate the representation of the OCEL log schema derived from our method, we used Coloured Petri nets (CPN), specifically CPN Tools\footnote{\url{https://cpntools.org/}}, to generate a simulated OCEL log for the cargo pickup process. 
The implementation details of the two CPN models and the simulated OCEL log are available from the project repository on GitHub\footnote{\url{https://github.com/jennyVVei/Dirigo_OCEL}}.

\section{Evaluation}
\label{sec:evaluation}
This section focuses on assessing the quality of OCEL log representations. 
We examine several existing object-centric log schemas using the port use case and evaluate the log representations using the proposed quality criteria.

\subsection{Existing Object-centric Log Schemas} 
\label{sec:existingocels}

As OCEL logs capture relationships between objects, current research~\citep{KumarST23} suggests they should be represented as databases rather than raw logs in a single file. \citet{KumarST23} compare two schemas for transforming OCEL logs to relational databases---STAR schema and normalised schema---and reveal that the normalised schema outperforms the STAR schema, as the latter tends to lose information about object relationships. Storing logs in a normalised form facilitates querying for integrity and compliance~\citep{KumarST23}. 
So far, there exist three different OCEL log representations in the form of a relational database.

\noindent 
\textbf{ACEL schema}~\citep{Moctar-MBabaASG22} is compliant with the OCEL 2.0 meta-model. Fig.~\ref{fig:acel} shows fragments of an event log following the ACEL schema. 
The representation of ACEL schema includes three relational tables, i.e., Events table, Objects table and Relations table. The Events table records all events executed in the process, associated objects and their dynamic attributes, O2O and E2O. The Objects table records the static attributes of all objects. The Relations table records all the relation sources and relation cardinality, which are static and mandatory, and the dynamic targets of the relations are recorded by the Events table.


\begin{figure*}[h]
\centering
\includegraphics[width=\textwidth]{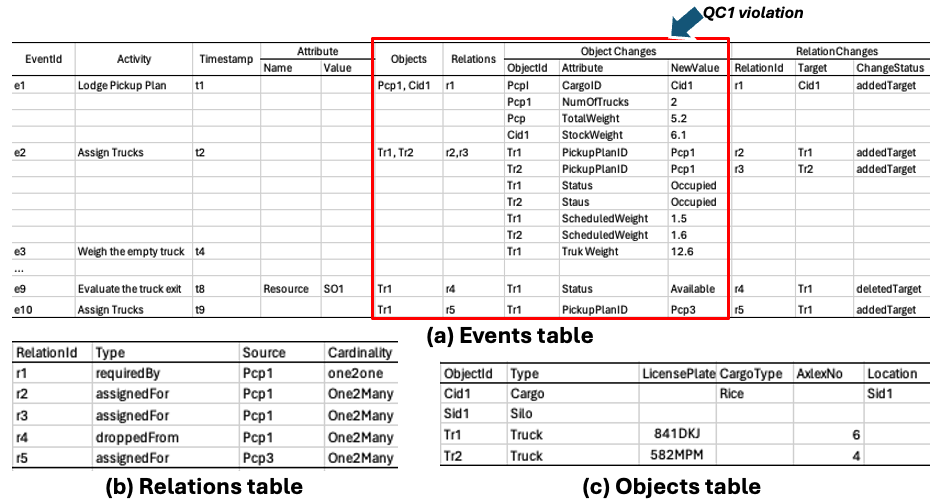}
\caption{Example of an event log following the ACEL schema~\citep{Moctar-MBabaASG22}} 
\label{fig:acel}
\end{figure*}

\noindent 
\textbf{DOCEL schema}~\citep{GoossensSVA22} partially conforms to the OCEL 2.0 meta-model. It consists of three types of relational tables: an Events table containing static event attributes, Object tables with static attributes, and Object tables with dynamic attributes. Fig.~\ref{fig:docel} illustrates fragments of an event log in DOCEL schema.

\begin{figure*}[h]
\centering
\includegraphics[width=\textwidth]{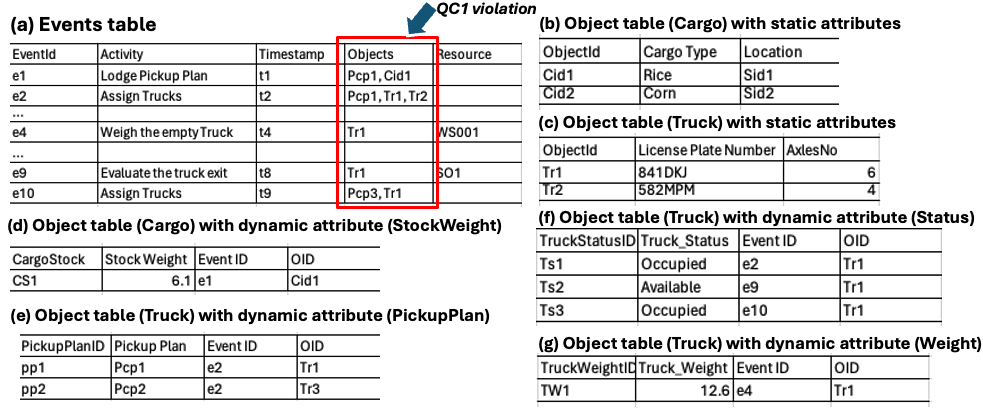} 
\caption{Example of OCEL log following the DOCEL schema~\citep{GoossensSVA22}} \label{fig:docel}
\end{figure*}

\noindent 
\textbf{XOC schema}~\citep{LiMCA18} includes a single table that details a sequence of events, capturing event types, object references, and object models (objects and their relations). Object attributes are not included in the XOC log representation. 
Fig.~\ref{fig:xoc} depicts an example fragment of an XOC log using the port use case.

\begin{figure*}[h!]
\centering
\includegraphics[width=\textwidth]{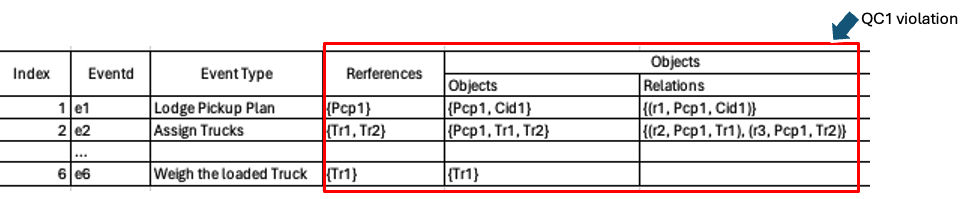}
\caption{Example of OCEL log following the XOC log Schema~\citep{LiMCA18}}
\label{fig:xoc}
\end{figure*}

\subsection{Evaluation of Object-centric Log Representations} 
In this section, we evaluate the quality of log representations of existing OCEL log schema (Section~\ref{sec:existingocels}) and the one derived using the \textit{Dirigo} method, based on the quality criteria outlined in Section~\ref{sec:qc}. We run various data queries for each quality criterion to demonstrate the evaluation results. The details of these queries are available in the project repository on GitHub. Table~\ref{tab:summaryqcs} summarises the evaluation results of different OCEL log representations according to the quality criteria.
\begin{table*}[h!!!]
\centering
\resizebox{.85\textwidth}{!}{%
\begin{tabular}{l|c|c|c|c}
\hline
\textbf{Quality Criteria} & \textbf{ACEL}             & \textbf{DOCEL}            & \textbf{XOC} & \textbf{Dirigo}           \\ \hline
QC1 (3rd NF Compliant)                                                                &                           &                           &              & \checkmark \\ \hline
QC2.a (No missing static object attributes)                                                              & \checkmark & \checkmark &              & \checkmark \\ \hline
QC2.b (No missing dynamic object attributes)                                                              &                           &                           &              & \checkmark \\ \hline
QC2.c (No missing event attributes)                                                              & \checkmark & \checkmark &              & \checkmark \\ \hline
QC3.a (No missing static O2O relations)                                                              & \checkmark & \checkmark &              & \checkmark \\ \hline
QC3.b (No missing dynamic O2O relations)                                                              & \checkmark & \checkmark &              & \checkmark \\ \hline
QC3.c (No missing E2O relations)                                                             & \checkmark & \checkmark &              & \checkmark \\ \hline
QC4.a (No missing labelled qualifier for O2O relations)                                                              & \checkmark &                           &              & \checkmark \\ \hline
QC4.b (No missing labelled qualifier for E2O relations)                                                              & \checkmark & \checkmark &              & \checkmark \\ \hline
\end{tabular}%
}
\caption{Summary of quality evaluation results}
\label{tab:summaryqcs}
\end{table*}

\paragraph{\textbf{QC1 (3rd NF Compliant)}} \mbox{} \\[0.5em]
\textbf{Clarification}: Except for \textbf{Dirigo Schema (Fig.~\ref{fig:dirigo_schema})}, all other three log presentations violate QC1.
\begin{itemize}[]
    \item \textbf{ACEL Schema (Fig.~\ref{fig:acel}a):}
    \begin{itemize}[]
        \item \textbf{Violation of 1st Normal Form:} The \texttt{objects} column contains multiple values.
        \item \textbf{Violation of 3rd Normal Form:} The \texttt{Attribute} column under the \texttt{ObjectChanges} column depends on the \texttt{ObjectID} column, creating a transitive dependency in the table (as \texttt{Attribute} depends on \texttt{ObjectID},
        and the latter is a non-key attribute depending on the key of the
        table consisting of the column \texttt{EventId}).
    \end{itemize}
    \item \textbf{DOCEL Schema (Fig.~\ref{fig:docel}a):}
    \begin{itemize}[]
        \item \textbf{Violation of 1st Normal Form:} The \texttt{Objects} column contains multiple values.
    \end{itemize}
    \item \textbf{XOC Schema (Fig.~\ref{fig:xoc}):}
    \begin{itemize}[]
        \item \textbf{Violation of 1st Normal Form:} Multiple values are present in the \texttt{References}, \texttt{Objects}, and \texttt{Relations} columns.
    \end{itemize}
\end{itemize}

\paragraph{\textbf{QC2.a (No missing static object attributes)}} \mbox{} \\[0.5em]
\noindent
\textbf{Example query}: \textit{What is the type of cargo Cid1?} Cargo type is a static attribute and remains unchanged in the cargo pickup process. 

\noindent
\textbf{Clarification}: Only \textbf{XOC Schema (Fig.~\ref{fig:xoc})} cannot answer this query as it does not include object information.
\begin{itemize}[]
    \item \textbf{Dirigo Schema (Fig.~\ref{fig:dirigo_schema}d)}: Contains both static and dynamic attributes of cargo, where we can identify the answer to this query is `Rice'.
    \item \textbf{ACEL Schema (Fig.~\ref{fig:acel}c)}: Records all static attributes of relevant objects, allowing us to identify the type of cargo Cid1 as `Rice'.
    \item \textbf{DOCEL Schema (Fig.~\ref{fig:docel}b)}: Captures static cargo attributes, thus also capable of answering the query.
\end{itemize}

\paragraph{\textbf{QC2.b (No missing dynamic object attributes)}} \mbox{} \\[0.5em]
\noindent
\textbf{Example query}: \textit{What is the last but one stock weight of cargo Cid1}? Cargo stock weight is a dynamic object attribute that is updated with each pickup. The log must record every change in value.

\noindent
\textbf{Clarification}: Only \textbf{Dirigo Schema (Fig.~\ref{fig:dirigo_schema})} can answer this query. By examining the \textbf{Events} table, \textbf{Object\_Cargo} table, and \textbf{E2O} table, we can determine the last but one stock weight of `Cid1'. None of the other three log representations can provide the answer to this query:
\begin{itemize}[]
    \item \textbf{ACEL Schema (Fig.~\ref{fig:acel}a)}: Records object changes, showing the latest stock weight of `Cid1' as `6.1' kg after lodging `Pcp1'. However, it does not record cargo stock history.
    \item \textbf{DOCEL Schema (Fig.~\ref{fig:docel}d)}: Stores dynamic attributes of Object cargo. It indicates a stock weight of `6.1' kg for `Cid1' at timestamp `t1' during the `Lodge pickup plan'. It is unable to identify the latest cargo stock before the pickup occurrence as every update of a dynamic attribute
    has to be associated with an event, so updates not considered to be events in the process can not be accessed.
    \item \textbf{XOC Schema (Fig.~\ref{fig:xoc})}: Lacks object information, making it incapable of answering this query.
\end{itemize}

\paragraph{\textbf{QC2.c (No missing event attributes)}} \mbox{} \\[0.5em]
\noindent
\textbf{Example query}: \textit{When was the truck with licence plate number `841DKJ' weighed before loading cargo Cid1?} The timestamp represents an event attribute.

\noindent
\textbf{Clarification}: Only \textbf{XOC Schema (Fig.~\ref{fig:xoc})} cannot answer this query because it orders events by index instead of timestamp. The rest of the log representations can provide the answer to this query. According to the domain knowledge, 
the first weighing activity before loading cargo is \pname{weigh the empty truck}.
\begin{itemize}[]
    \item \textbf{Dirigo Schema (Fig.~\ref{fig:dirigo_schema})}: Includes the \textbf{Object\_Truck} table, \textbf{Events} table, and \textbf{E2O} table. By examining these three tables, we can obtain the answer (`t4') to this query.
    \item \textbf{ACEL Schema (Fig.~\ref{fig:acel}a \& Fig.~\ref{fig:acel}c)}: By examining the \textbf{Objects} table (for identify the object ID \pname{Tr1} for the truck with the license plate number `841DKJ) in the ACEL schema and the \textbf{Events} table, we obtain timestamp `t4'. 
    \item \textbf{DOCEL Schema (Fig.~\ref{fig:docel}a \& Fig.~\ref{fig:docel}c)}: By referring to the \textbf{Events} and the \textbf{Objects} tables for \texttt{objectId} (`Tr1'), we obtain timestamp `t4'.
\end{itemize}


\paragraph{\textbf{\textit{QC3.a (No missing static O2O relations)}}} \mbox{} \\[0.5em]
\noindent
\textbf{Example Query}: \textit{Which silo in the port is cargo Cid1 stored in?} In the cargo pickup process, the relationship between cargo and silo is static because changing the location of cargo storage is outside the scope of this pickup process. Thus, we assume this relationship remains constant.

\noindent
\textbf{Clarification}: The \textbf{XOC Schema (Fig.~\ref{fig:xoc})} yields a null result, as it does not include object information. All the other three log representations can address this query. 
\begin{itemize}[]
    \item \textbf{Dirigo Schema (Fig.~\ref{fig:dirigo_schema}d)}: Provides the information that cargo `Cid1' is stored in `Sid1'.
    \item \textbf{ACEL Schema (Fig.~\ref{fig:acel}c)}: Captures the answer by referring to \texttt{objectId} `Cid1' and attribute \texttt{Location} (yielding the value `Sid1').
    \item \textbf{DOCEL Schema (Fig.~\ref{fig:docel}b)}: Contains cargo static attributes, showing the stored location is `Sid1'.
\end{itemize}


\paragraph{\textbf{\textit{QC3.b (No missing dynamic O2O relations)}}} \mbox{} \\[0.5em]
\noindent
\textbf{Example Query}: \textit{To which pickup plan was the truck with licence plate number `841DKJ' assigned after completing its immediately preceding pickup plan?} An example of dynamic O2O relations in the cargo pickup process is the relationship between trucks and pickup plans. Trucks may be reassigned to a different pickup plan after completing the current one, thus altering their relationship.

\noindent
\textbf{Clarification}: Only \textbf{XOC Schema (Fig.~\ref{fig:xoc})} fails to answer this query. Although the XOC log captures \pname{Assign Trucks} events and associated objects, it lacks specific truck details, such as licence plate numbers.
\begin{itemize}[]
    \item \textbf{Dirigo Schema (Fig.~\ref{fig:dirigo_schema})}: By examining the \textbf{Object\_Truck} table and the \textbf{O2O} table, we can identify that a truck with \texttt{object\_id} (`Tr1') gets dropped from pickup plan `Pcp1' and immediately assigned to pickup plan ‘Pcp3’.
    \item \textbf{ACEL Schema (Fig.~\ref{fig:acel})}: Can resolve this query through the \textbf{Events}, \textbf{Objects}, and \textbf{Relations} tables. By mapping the license plate number to the corresponding object \pname{Tr1} and by joining the \textbf{Events} and \textbf{Relations} tables through \texttt{RelationId}, we determine the truck's initial pickup plan as `Pcp1' and the subsequent reassignment to `Pcp3'.
    \item \textbf{DOCEL Schema (Fig.~\ref{fig:docel}a \& Fig.~\ref{fig:docel}c)}: Provides the answer through the \textbf{Objects} table, capturing the static attribute of trucks and the \textbf{Events} table. We find the reassignment details by identifying the object ID for the truck with licence plate number `841DKJ' and examining the Events table for `Assign Trucks' activities related to \texttt{objectId} `Tr1'.
\end{itemize}

\paragraph{\textbf{\textit{QC3.c (No missing E2O relations)}}} \mbox{} \\[0.5em]
\noindent
\textbf{Example Query} \textit{What is the weight of the truck with licence plate number ``841DKJ'' before loading cargo Cid1?} Truck weight is a dynamic attribute that changes with different weighing activities. To determine a truck's weight before loading cargo, we must identify the associated \pname{Weigh the empty truck} activity.
   
\noindent
\textbf{Clarification}: The \textbf{XOC log (Fig.~\ref{fig:xoc})} captures events with the activity `Weigh the empty truck' and associated objects but lacks specific truck information, such as license plate numbers, thus cannot answer this query. All other three log representations can answer this query.
\begin{itemize}[]
    \item \textbf{Dirigo Schema (Fig.~\ref{fig:dirigo_schema}a, Fig.~\ref{fig:dirigo_schema}b, Fig.~\ref{fig:dirigo_schema}c \& Fig.~\ref{fig:dirigo_schema}e)}: Provides the answer by joining tables \textbf{E2O}, \textbf{Events}, \textbf{O2O} and \textbf{Object\_Truck} through \texttt{Timestamp}. By identifying the truck with the specified license plate number and the time it was assigned to pick up cargo `Cid1', we can determine the time of weighing the empty truck.
    \item \textbf{ACEL Schema (Fig.~\ref{fig:acel})}: Can answer this query by referring to the \textbf{Objects} and \textbf{Events} tables. First, we identify the \texttt{objectId} for the truck (`Tr1') based on the license plate number. Next, we use this \texttt{objectId} to look up
    the row where its \texttt{Activity name} is `Weigh the Empty Truck' in the \textbf{Events} table, determining the truck's weight as `12.6'.
    \item \textbf{DOCEL Schema (Fig.~\ref{fig:docel}a \& Fig.~\ref{fig:docel}g)}: Can provide the answer to this query by first identifying the \texttt{objectId} of the truck based on the license plate number. Then, we use this to identify the \texttt{EventId} with the activity name `Weigh the Empty Truck' in the \textbf{Events} table for that truck. By cross-referencing the Event ID and Object ID in the \textbf{Object} table, the truck's weight can be found. 
\end{itemize}

\paragraph{\textbf{\textit{QC4.a (No missing labelled qualifier for O2O relations)}}} \mbox{} \\[0.5em]
\noindent
\textbf{Example Query} \textit{When was the truck with license plate number `841DKJ' dropped from its immediately preceding pickup plan?}   In object-centric processes, object relationships, such as the assignment of trucks to pickup plans, can change. It is essential to use qualifier labels to differentiate these O2O relations.

\noindent
\textbf{Clarification}: \textbf{DOCEL Schema} and \textbf{XOC Schema} cannot answer this query.
\begin{itemize}[]
    \item \textbf{Dirigo Schema (Fig.~\ref{fig:dirigo_schema}c \& Fig.~\ref{fig:dirigo_schema}e)}: Can answer this query by retrieving the truck with the specified license plate number in the \textbf{Object\_Truck} table and then directly referring to the \textbf{O2O} table.
    \item \textbf{ACEL Schema (Fig.~\ref{fig:acel}a \& Fig.~\ref{fig:acel}c)}: Can provide the answer to this query. Using the \textbf{Objects} table, we identify the object ID \pname{Tr1} for the truck with the license plate number `841DKJ'. The \textbf{Events} table shows that in event `e2' at timestamp `t2', with the activity `Assign Truck', `Tr1' was assigned to pickup plan `Pcp1'. This is indicated by `r2' in the \texttt{RelationId} column and `addedTarget' in the \texttt{ChangeStatus} column, mapping to the relation in the \textbf{Relations} table where the source is `Pcp1'. At timestamp `t8', in event `e9', `Tr1' is involved again, with the \texttt{RelationId} `r4' and the \texttt{ChangeStatus} value `deletedTarget'. Querying the \textbf{Relations} table using `r4' reveals that `Tr1' was dropped from `Pcp1'. Thus, we can conclude that the truck with license plate number `841DKJ' was dropped from its pickup plan `Pcp1' during event `e9' at timestamp `t8'.
    \item \textbf{DOCEL Schema (Fig.~\ref{fig:docel})}: Cannot answer this query as it does not include any semantics of relationships between objects~\citep{GoossensSVA22}.
    \item \textbf{XOC Schema (Fig.~\ref{fig:xoc})}: Cannot answer this query because it lacks specific truck details. The object IDs do not indicate which truck corresponds to license plate number `841DKJ'. While the schema captures E2O and O2O relations, these are labelled generically (e.g., `r1' and `r2') without semantic clarity. Consequently, the relationship meanings between objects, such as `Pcp1' and `Tr1', remain ambiguous.
\end{itemize}

\paragraph{\textbf{\textit{QC4.b (No missing labelled qualifier for E2O relations)}}} \mbox{} \\[0.5em]
\noindent
\textbf{Example Query} \textit{Show all the times at which the truck with licence plate number `841DKJ' changed its status from occupied to available.} The semantics of E2O relations indicate the role of objects in events. In the cargo pickup process, trucks are assigned after lodging the pickup plan and dropped from it after completion. By tracking changes in the dynamic truck attribute (i.e., status), we can understand the roles trucks play in different activities within the process.

\noindent
\textbf{Clarification}: Only \textbf{XOC Schema (Fig.~\ref{fig:xoc})} cannot answer this query as the XOC log does not include any semantics of E2O relations.
\begin{itemize}[]
    \item \textbf{Dirigo Schema (Fig.~\ref{fig:dirigo_schema}e)}: Can answer this query through the \textbf{Object\_Truck} table.
    \item \textbf{ACEL Schema (Fig.~\ref{fig:acel}a \& Fig.~\ref{fig:acel}c)}: Can address this query. Using the \textbf{Objects} table, we identify the object ID \pname{Tr1} for the truck with the license plate number `841DKJ'. We then search the \texttt{ObjectChanges} column in the \textbf{Events} table for entries where the \texttt{ObjectId} is `Tr1', and the \texttt{Attribute} is `Status'. In particular, we look for entries where the `Status' value changes from `Occupied' to `Available'. The \textbf{Events} table shows that at timestamp `t2', `Tr1's status was `Occupied', and at timestamp `t8', it changed to `Available'. Thus, the truck with license plate number `841DKJ' (`Tr1') changed its status from occupied to available at timestamp `t8'.

    \item \textbf{DOCEL Schema (Fig.~\ref{fig:docel}a, Fig.~\ref{fig:docel}b \& Fig.~\ref{fig:docel}f)}: Can provide the answer to this query. First, we identify the truck's object ID from the table, capturing the truck's static attributes. We then track the truck's status changes using the table capturing its dynamic attribute---status. Truck `Tr1' was `Occupied' at event `e2, `Available' at event `e9', and `Occupied' again at event `e10'. By identifying the timestamps for these events using the \textbf{Events} table, we obtain the answer for this query.
\end{itemize}


\subsection{Discussion}

In addition to the above evaluation, it is worthy noting that:


\begin{itemize}[]
    \item While \textbf{ACEL} largely aligns with the OCEL 2.0 meta-model, its log representation quality suffers due to excessive information, particularly because of the presence of multiple hierarchical levels of columns within the \textbf{Events} table. This violates the third normal form, leading to significant data redundancy.
    \item \textbf{DOCEL} stores each dynamic attribute of an object in a separate table. This schema assigns an identifier to each dynamic attribute and uses two foreign keys—the Event ID and the Object ID—allowing changes to the attribute and its value to be traced back to the corresponding object and the event associated with it~\citep{GoossensSVA22}. However, this significantly increases the log complexity.
    \item \textbf{DOCEL} integrates a dynamic object attribute with Event ID and Object ID in a single table.  By associating an object's dynamic attributes with the corresponding events, DOCEL can capture the precise relationship between an object and an event, avoiding the ambiguity that arises when an object’s dynamic attributes correspond to different events but share the same timestamp. However, when an event is associated with multiple dynamic attributes of an object, the DOCEL tables capturing these attributes will duplicate the Event ID and Object ID. For instance, the relationship between `e2' and `Tr1' is duplicated in more than one table, capturing the dynamic attributes of the object `Tr1' (as illustrated in Fig.~\ref{fig:docel}e \& Fig.~\ref{fig:docel}f).

\end{itemize}
\section{Conclusion}
\label{sec:conclusion}
In this work, we have proposed a method, \textit{Dirigo}, to systematically extract event logs for object-centric processes, as well as a set of quality criteria to assess the quality of OCEL log representations. The design of the proposed method is guided by the established quality criteria and design principles (conceptualisation principle, 100\% principle and being generic).
To validate our method, we have applied \textit{Dirigo} to a real-life use case, extracting an OCEL log via simulation. We then assess the quality of the log representation of the extracted event log in comparison to those of existing OCEL logs, employing the established quality criteria. 

Our evaluation demonstrates that the OCEL log extracted following the \textit{Dirigo} method, exhibits high-quality log representation compared to existing object-centric log schemas, according to the proposed quality criteria. 
The OCEL log schema extracted via \textit{Dirigo} adheres to the OCEL 2.0 meta-schema and extends it by including temporal information of dynamic object relationships and explicitly capturing resources as event attributes. Future work will focus on object-centric process analytics, exploring how the event logs extracted using \textit{Dirigo} can be used to enhance object-centric process analysis.

\section*{Statements and Declarations}
The authors have no competing interests to declare that are relevant to the content of this article.

\bibliography{sn-bibliography}

\end{document}